%Paper: hep-ph/9409357
%From: jfgucd@ucdhep.ucdavis.edu
%Date: Mon, 19 Sep 1994 18:00:18 PST

\input tables.tex
\input phyzzx.tex
\def\yh{y_{\h}}

\def\vevlam{\VEV{\lam\lam^\prime}}
\def\eepem{E_{\epem}}

\def\ie{{\it i.e.}}
\def\epem{e^+e^-}

\def\chitil{\widetilde\chi}
\def\mgluino{m_{\widetilde g}}
\def\mt{m_t}

\def\mgut{M_{U}}

\def\dil{D}
\def\sdil{SD}

\def\dilp{\dil^+}
\def\dilm{\dil^-}

\def\sdilp{\sdil^+}
\def\sdilm{\sdil^-}

\def\mgltb{$\mgl\,$--$\tanb$}
\def\hl{h^0}
\def\hh{H^0}
\def\ha{A^0}
\def\hm{H^-}
\def\hp{H^+}
\def\mhl{m_{\hl}}
\def\mhh{m_{\hh}}
\def\mha{m_{\ha}}
\def\hpm{H^{\pm}}
\def\mhpm{m_{\hpm}}
\def\hp{H^+}
\def\gam{\gamma}

\def\gev{~{\rm GeV}}

\def\fbi{~{\rm fb}^{-1}}

%%Journal definitions

\def\prdj#1{{\it Phys. Rev.} {\bf D{#1}}}

\def\prlj#1{{\it Phys. Rev. Lett.} {\bf {#1}}}
\def\plbj#1{{\it Phys. Lett.} {\bf B{#1}}}

\def\mt{m_t}
\def\wp{W^+}
\def\wm{W^-}

\def\rta{\rightarrow}

\def\anti{\overline}

\def\fbi{~{\rm fb}^{-1}}

\def\VEV#1{\left\langle #1\right\rangle}
\catcode`\@=11 % This allows us to modify PLAIN macros.

\def\t1{{\tilde 1}}

\def\etmiss{\slash {E_T}}

\def\gev{\,{\rm GeV}}

\def\wt{\widetilde}

\def\rta{\rightarrow}

\def\gl{\wt g}
\def\mgl{m_{\gl}}
\def\stop{\wt t}

\def\mstop{m_{\stop}}

\def\sq{\wt q}

\def\msq{m_{\sq}}
\def\slep{\wt l}

\def\slepl{\slep_L}
\def\slepr{\slep_R}
\def\mslepl{m_{\slepl}}
\def\mslepr{m_{\slepr}}

\def\stopi{\wt t_1}
\def\mstopi{m_{\stopi}}

\def\hl{h^0}
\def\hh{H^0}
\def\ha{A^0}
\def\hp{H^+}
\def\hm{H^-}
\def\hpm{H^{\pm}}
\def\mhl{m_{\hl}}
\def\mhh{m_{\hh}}
\def\mha{m_{\ha}}

\def\tanb{\tan\beta}
\def\mt{m_t}

\def\mgut{M_U}
\def\mstring{M_S}
\def\wp{W^+}
\def\wm{W^-}

\def\cnone{\wt\chi^0_1}
\def\cntwo{\wt\chi^0_2}
\def\snu{\wt\nu}

\def\msnu{m_{\snu}}
\def\mcnone{m_{\cnone}}
\def\mcntwo{m_{\cntwo}}
\def\h{h}
\def\cpone{\wt \chi^+_1}
\def\cmone{\wt \chi^-_1}
\def\cpmone{\wt \chi^{\pm}_1}
\def\mcpone{m_{\cpone}}

\def\tanb{\tan\beta}

\def\nsd{N_{SD}}

\def\mw{m_W}

\def\anti{\overline}

\def\ifmath#1{\relax\ifmmode #1\else $#1$\fi}

\def\3quarter{{\textstyle{3 \over 4}}}

\def\h{h}
\def\mh{m_{\h}}

\def\eepem{E_{\epem}}

\def\hpm{H^{\pm}}
\def\mhpm{m_{\hpm}}

\def\hp{H^+}
\def\hm{H^-}

\def\chitil{\widetilde\chi}
\def\cnone{\chitil^0_1}

\def\h{h}
\def\mh{m_{\h}}
\def\tanb{\tan\beta}

\def\chitil{\widetilde\chi}

\def\cnone{\chitil^0_1}
\def\cntwo{\chitil^0_2}

\def\mcnone{m_{\chitil^0_1}}
\def\mcntwo{m_{\chitil^0_2}}

\def\wp{W^+}
\def\wm{W^-}
\def\hl{h^0}
\def\mhl{m_{\hl}}
\def\hh{H^0}
\def\mhh{m_{\hh}}
\def\hpm{H^{\pm}}
\def\mhpm{m_{\hpm}}
\def\mt{m_t}

\def\hl{h^0}
\def\hh{H^0}
\def\ha{A^0}
\def\mhl{m_{\hl}}
\def\mhh{m_{\hh}}
\def\mha{m_{\ha}}

\def\lam{\lambda}

\def\gam{\gamma}
\def\gam{\gamma}

\def\gam{\gamma}

\def\stop{{\wtilde t}}
\def\mstop{m_{\stop}}
\def\ep{e^+}
\def\em{e^-}

\def\etmiss{E_T^{\,miss}}

\def\gev{~{\rm GeV}}

\def\fbi{~{\rm fb}^{-1}}

\def\wt{\widetilde}

%%Journal definitions

\def\prdj#1{{\it Phys. Rev.} {\bf D{#1}}}

\def\prlj#1{{\it Phys. Rev. Lett.} {\bf {#1}}}
\def\plbj#1{{\it Phys. Lett.} {\bf B{#1}}}

\def\mt{m_t}
\def\wp{W^+}
\def\wm{W^-}
\def\rta{\rightarrow}
\def\tanb{\tan\beta}

\def\nsd{N_{SD}}

\def\mw{m_W}

\def\anti{\overline}

\def\ifmath#1{\relax\ifmmode #1\else $#1$\fi}

\def\3quarter{{\textstyle{3 \over 4}}}

\input phyzzx
\Pubnum={$\caps UCD-94-35$\cr}
\date{September 1994}

\titlepage
\vskip 0.75in
\baselineskip 0pt %%%%%%%%%physrev -- remove
%\PHYSREV %%%%%%%%%%physrev -- include
\hsize=6.5in
\vsize=8.5in
\centerline{{\bf Detection of Minimal Supersymmetric Model Higgs Bosons}}
\centerline{{\bf in $\gam\gam$ Collisions: Influence of SUSY Decay Modes}}
\vskip .075in
\centerline{ J.F. Gunion, J.G. Kelly, and J. Ohnemus}
\vskip .075in
\centerline{\it Davis Institute for High Energy Physics,
Dept. of Physics, U.C. Davis, Davis, CA 95616}

\vskip .075in
\centerline{\bf Abstract}
\vskip .075in
\centerline{\Tenpoint\baselineskip=12pt%\parindent=1pc
\vbox{\hsize=12.4cm
\noindent We demonstrate that supersymmetric decay modes
of the neutral Higgs bosons of the MSSM could well make their
detection extremely difficult when produced singly in $\gam\gam$ collisions
at a back-scattered laser beam facility.
}}

\vskip .15in
\noindent{\bf 1. Introduction}
\vskip .075in

Supersymmetric models are leading candidates for extending the Standard Model
(SM).
\REF\hhg{For a review see
J.F. Gunion, H.E. Haber, G.L. Kane, and S. Dawson, {\it The Higgs Hunter's
Guide}, Addison-Wesley, Redwood City, CA (1990).}\
The simplest such model is the minimal supersymmetric model (MSSM),
which is defined by having precisely two Higgs doublets.\refmark{\hhg}\
The physical Higgs eigenstates comprise two charged Higgs
bosons ($\hpm$), two CP-even Higgs bosons ($\hl$ and $\hh$ with $\mhl<\mhh$),
and one CP-odd Higgs boson ($\ha$).
\REF\habgun{J.F. Gunion and H.E. Haber, Proceedings of the 1990 DPF
Summer Study on High Energy Physics, ``Research Directions
for the Decade'', Snowmass (1990), ed. E. Berger, p. 469; \prdj{48}
(1993) 2907.}
\REF\caldwelletal{D. Borden, D. Bauer, and D. Caldwell, \prdj{48} (1993)
4018. See also D. Borden, Proceedings of the ``Workshop on Physics
and Experiments with Linear $\epem$ Colliders'', Waikaloa, Hawaii,
April 26-30 (1993), eds. F.A. Harris, S.L. Olsen, S. Pakvasa, and X. Tata,
p. 323.}
A possibly very important means for discovering the neutral MSSM Higgs
bosons at an $\epem$ collider is to produce them via collisions of
polarized photons\refmark{\habgun,\caldwelletal}\
obtained by back-scattering polarized laser beams off of
polarized electron and positron beams at a TeV-scale linear $\ep\em$ collider.
\REF\telnovi{H.F. Ginzburg, G.L. Kotkin, V.G. Serbo, and V.I. Telnov,
{\it Nucl. Inst. and Meth.} {\bf 205} (1983) 47.}
\REF\telnovii{H.F. Ginzburg, G.L. Kotkin, S.L. Panfil,
V.G. Serbo, and V.I. Telnov,
{\it Nucl. Inst. and Meth.} {\bf 219} (1984) 5.}
\refmark{\telnovi,\telnovii}\ In previous work,\refmark{\habgun}\
it has been established that the neutral MSSM Higgs bosons can indeed
be detected in $\gam\gam$ collisions over much of parameter space,
{\it provided they decay primarily to SM final states}.
In fact, since the possibly heavy $\hh$ and $\ha$ can be produced singly
by direct
$\gam\gam$ collisions, whereas they are only detectable in $\epem$ collisions
in the pair production mode,  $\epem\rta\ha\hh$, photon-photon colliders can
even provide a larger discovery mass reach than
direct $\epem$ collisions.\refmark\habgun\
However, an open question is the extent to which the possibilities
for $\ha$ and $\hh$ detection in $\gam\gam$ collisions are altered by
significant decays to supersymmetric particle channels.
In this paper, we show that such decays could have a decidedly negative
impact.

The importance of supersymmetric decays of the MSSM Higgs bosons is
dictated by the parameters of soft supersymmetry breaking.
The four basic parameters are:
a) the gaugino masses $M_a$ (where $a$ labels the group);
b) the scalar masses $m_i$ (where $i$ labels the various
scalars, \eg\ Higgs bosons, sleptons, squarks);
c) the soft Yukawa coefficients $A_{ijk}$; and
d) the $B$ parameter which specifies the
soft mixing term between the two Higgs scalar fields.
The success of gauge coupling unification in the context of the MSSM
lends considerable credence not only to the
possibility that this extension of the Standard Model
is correct, but also to the idea that the boundary conditions
for all the soft-supersymmetry-breaking parameters at
the unification scale could be relatively simple and universal.
Superstring theory provides particularly
attractive and well-motivated examples of such boundary conditions.
In this paper we consider
the dilaton-like superstring supersymmetry-breaking scheme
(labelled by D). This is one of the most attractive models available and
yields a complex array of decay channels for the MSSM Higgs bosons.
In this model the $M_a$, $m_i$, and $A_{ijk}$ parameters all take on
universal values at the unification scale $\mgut$ related by:
$$
M^0=-A^0=\sqrt3m^0.
\eqn\bc$$
Predictions in this model for the $B$ parameter
are rather uncertain, and so it is kept a free parameter.
The dilaton-like boundary conditions are certainly those appropriate
when supersymmetry breaking is dominated by the dilaton field in
string theory, but they also apply for a remarkably broad class of models
(including Calabi-Yau compactifications, and orbifold models in which
the MSSM fields all belong to the untwisted sector) so long
as the moduli fields do not play a dominant role in supersymmetry breaking.
\REF\bgkp{H. Baer, J.F. Gunion, C. Kao, and H. Pois, preprint UCD-94-19.}
For a brief review and detailed references, see Ref.~[\bgkp].

If the boundary conditions of Eq.~\bc\ are imposed
and the top quark mass is fixed [we
adopt $\mt(\mt)=170\gev$, corresponding to a pole mass of
about 178 GeV] only two free parameters and a sign remain undetermined
after minimizing the potential. The two parameters can be taken to be
$\tanb$, the ratio of the neutral Higgs field vacuum expectation values, and
$\mgluino$, the gluino mass. The parameter $B$ is determined in terms of
these, as are all other superpotential parameters, including the magnitude of
the Higgs superfield mixing parameter $\mu$. However, the sign of $\mu$ is
not determined. Two models result --- $\dilp$ and
$\dilm$, the superscript indicating the sign of $\mu$ --- the phenomenology
of which can be explored in the two dimensional \mgltb\ parameter space.

The discussion so far has obscured one fundamental problem facing
the gauge coupling unification success: namely, the scale  $\mgut$
at which the couplings naturally unify is $\sim 2\times 10^{16}\gev$,
\ie\ much less than the natural scale for supergravity and string unification
of
$\mstring\sim10^{18}\gev$.  A variety of excuses for this have been
discussed. In Ref.~[\bgkp] two extreme approaches were adopted:
i) ignore the difference ---
a more complete understanding of the feed-down of SUSY breaking from the
full supergravity or superstring theory could resolve the discrepancy;
ii) assume that the unification at $\mgut$ is only apparent (\ie\ accidental)
and introduce a minimal set of additional matter fields at high scale
with masses chosen precisely so as to give coupling unification
at $\mstring$.  We will not go into detail regarding these extra fields;
a discussion and references can be found in Ref.~[\bgkp].
The models with such extra fields are termed the `string-scale-unified'
versions of the previously listed models, and will be denoted by
$\sdilp$ and $\sdilm$.

To systematically investigate the resulting models, Ref.~[\bgkp] first
established the allowed region of \mgltb\ parameter space for each subject to:
a) all predicted SUSY partner particles (including the light Higgs boson
$\hl$) are unobservable;
b) the lightest SUSY particle is either the lightest neutralino
$\cnone$ (as is always the case for the allowed parameter space
of the models explored here) or the sneutrino $\snu$;
c) the top quark Yukawa coupling remains perturbative at all scales
from $\mw$ up to $\mgut$ or $\mstring$; and
d) proper electroweak symmetry breaking and a global minimum are obtained.
Constraints from $b\rta s\gamma$,
relic abundance, and proton decay were not imposed, as these all have
considerable uncertainties and/or require additional model-dependent input.
Exact $b-\tau$ Yukawa unification was also not imposed.

Within the allowed parameter spaces, the masses of the
SUSY particles scale with $\mgluino$; variation of the masses with $\tanb$
at fixed $\mgluino$ is relatively limited, especially for $\mgluino$
values above about 500 to 600 GeV,
with $\slepr,\cpone,\cntwo,\snu,\slepl$ clustering between 0.2 to 0.4 times
$\mgluino$. It is the restricted size of the soft scalar
mass parameter, $m^0$, relative to $M^0$ that causes the sleptons to be
rather light in the dilaton-like models.
Indeed, slepton masses are largely generated by renormalization-group
evolution from the $M^0$ gaugino seed value at $\mgut$; only the squarks
acquire masses comparable to $\mgluino$, as a result of the driving terms
proportional to $\alpha_s$ in the RGE's.

Regarding the Higgs boson masses, a very general pattern emerges.
The $\hl$ is normally relatively light, even after
including the standard
one-loop radiative corrections,\Ref\radcorra{H. Haber and R. Hempfling
\prlj{66} (1991) 1815; Y. Okada, M.
Yamaguchi, and T. Yanagida, {\it Prog. Theor. Phys.} {\bf 85} (1991) 1;
J. Ellis, G. Ridolfi, and F. Zwirner, \plbj{257} (1991) 83.}\
which depend most crucially upon the
top quark mass ($\mt$) and the stop squark mass ($\mstop$).
For gluino masses below 1 TeV and $\mt(\mt)=170$ GeV, $\mhl\leq 125\gev$,
with quite low values ($65\lsim\mhl\lsim 110\gev$) being rather typical.
Thus, the $\hl$ will be easily discovered via $\epem\rta Z\hl$
(even if the $\hl$ decays invisibly to $\cnone\cnone$,
as can happen in these models).  In contrast, the RGE driven electroweak
symmetry breaking models in general, and the dilaton-like boundary
condition models in particular, predict rather large $\mha
\sim \mhh\sim\mhpm$ values. For most of parameter space,
$\mha\gsim 200\gev$ with values in the $300-600\gev$ range being
much more typical for $\mgluino<800\gev$.
\foot{This upper bound represents a purely aesthetical choice as to
an $\mgluino$ value below which the model is clearly not fine-tuned.}
This means that $\epem\rta \ha\hh,\hp\hm$ pair production
is quite possibly disallowed kinematically for a $\sqrt s \sim 500\gev$
$\epem$ collider, and that single production via $\gam\gam\rta \ha,\hh$
would be the only possible mode of discovery. Further, for
$\mgluino\lsim 800\gev$ the $\cnone$, $\cntwo$, $\cpmone$, $\snu$, $\slepr$,
and (except at high $\tanb$) $\slepl$ are all light enough to appear in
two-body
decay modes of the $\ha$ and $\hh$.  Thus, the $\dil$ and $\sdil$ models
present many possible scenarios of precisely the type that we wish to
explore.

\vskip .15in
\noindent{\bf 2. Scenarios}
\vskip .075in

\TABLE\scenarios{A tabulation of supersymmetric particle masses (in GeV)
for the $\dil$ and $\sdil$ scenarios considered.}

 \midinsert
 \titlestyle{\tenpoint
 Table \scenarios: A tabulation of supersymmetric particle masses (in GeV)
for the $\dil$ and $\sdil$ scenarios considered.
}
 \smallskip

 \thicksize=0pt
 \hrule \vskip .04in \hrule
 \begintable
 Scenario | $\mgl$ | $\tanb$ | $\mhl$ | $\mha$ | $\mhh$ |
$\mcnone$ | $\mcntwo$ |
$\mcpone$ | $\mslepl$ |   $\mslepr$ | $\msnu$ | $\msq$ | $\mstopi$ \cr
$\dilp_3$ | 310 | 15.0 | 103 |  180 | 180 |
39.9 | 72.5 | 70.2 | 109 | 85.9 | 74.4 | 277 | 188 \nr
$\dilm_1$ | 232 | 2.0  | 58.4 | 190 | 205 |
37.1 | 83.5 | 83.3 | 82.3 | 65.0 | 54.1 | 207 | 215 \nr
$\dilm_4$ | 301 | 2.2  | 69.0 | 244 | 255 |
47.3 | 100  | 100  | 103  | 80.2 | 79.8 | 269 | 242 \nr
$\dilp_4$ | 346 | 3.2  | 93.6 | 250 | 255 |
40.4 | 79.2 | 73.5 | 118  | 91.8 | 93.0 | 310 | 195 \nr
$\dilp_5$ | 431 | 4.5  | 104 | 300 | 302 |
58.4 | 109  | 107  | 144  | 111  | 122  | 386 | 250 \nr
$\dilp_7$ | 503 | 5.0  | 108 | 350 | 351 |
71.3 | 134  | 133  | 166  | 127  | 147  | 450 | 297 \nr
$\sdilm_1$ | 471 | 15.0 | 111  | 357 | 357 |
69.1 | 134  | 134  | 193  | 157  | 176  | 464 | 301 \nr
$\sdilm_2$ | 503 | 5.0  | 105  | 424 | 426 |
75.4 | 149  | 149  | 205  | 166  | 190  | 496 | 339
\endtable
 \hrule \vskip .04in \hrule
\vfill
 \endinsert

Of the specific \mgltb\ scenarios explored with regard to
their general phenomenology in Ref.~[\bgkp], we focus on a limited
number of representative cases. In the notation of Ref.~[\bgkp],
these are the scenarios $\dilp_3$, $\dilm_1$, $\dilm_4$, $\dilp_4$,
$\dilp_5$, $\dilp_7$, $\sdilm_1$, and $\sdilm_2$, where we have
listed them in order of increasing $\mha$.
A complete listing of all relevant
particle masses, and a summary of the decay modes of the SUSY particles
is presented in Ref.~[\bgkp].  Here, we give a condensed summary along
with details regarding the decays of the $\ha$ and $\hh$ Higgs bosons.
The scenarios are summarized in Table~\scenarios, where
we give masses for the Higgs bosons and selected superparticles.

\TABLE\brs{A tabulation of important branching ratios
for a) the $\hh$ and b) the $\ha$. In the results $\slep=\wt e,\wt\mu$
are summed together and all $\snu\,\snu$ modes are summed together.
We use the shorthand notation $\slep\ \slep\equiv \slepl\slepl+\slepr\slepr$.}

\midinsert
 \titlestyle{\tenpoint
Table \brs a: A tabulation of important branching ratios
for the $\hh$. In the results $\slep=\wt e,\wt\mu$
are summed together and all $\snu\,\snu$ modes are summed together.
We use the shorthand notation $\slep\ \slep\equiv \slepl\slepl+\slepr\slepr$.
}
 \smallskip

 \thicksize=0pt
 \hrule \vskip .04in \hrule
\begintable
 Scenario | $b\anti b$ | $t\anti t$ |
            $\wp\wm+ZZ$ | $\cnone\cnone$ | $\cnone\cntwo$ |
            $\cntwo\cntwo$ | $\cpone\cmone$ | $\hl\hl$ |
            $\slep\ \slep$ |  $\snu\,\snu$ \cr
$\dilp_3$ | 0.782 | --- | 0.0003 | 0.031 | 0.046 | 0.011 | 0.072 | --- |
            0.0003 | 0.003 \nr
$\dilm_1$ | 0.045 | --- | 0.038 | 0.002 | 0.031 | 0.088 | 0.112 | 0.103 |
            0.110 | 0.414 \nr
$\dilm_4$ | 0.072 | --- | 0.054 | 0.004 | 0.053 | 0.105 | 0.155 | 0.149 |
            0.081 | 0.280 \nr
$\dilp_4$ | 0.144 | --- | 0.038 | 0.104 | 0.126 | 0.034 | 0.292 | 0.064 |
            0.034 | 0.136 \nr
$\dilp_5$ | 0.343 | --- | 0.024 | 0.062 | 0.136 | 0.060 | 0.247 | 0.028 |
            0.014 | 0.054 \nr
$\dilp_7$ | 0.456 | 0.030 | 0.018 | 0.040 | 0.113 | 0.058 | 0.187 | 0.016 |
            0.009 | 0.032 \nr
$\sdilm_1$ | 0.833 | 0.001 | 0.002 | 0.005 | 0.022 | 0.019 | 0.042 | 0.013 |
             0.0001 | 0.0003 \nr
$\sdilm_2$ | 0.315 | 0.273 | 0.018 | 0.010 | 0.057 | 0.068 | 0.134 | 0.082 |
             0.004 | 0.013 %\nr
\endtable
 \hrule \vskip .04in \hrule
\bigskip
 \titlestyle{\tenpoint
 Table \brs b: A tabulation of important branching ratios for the $\ha$.
}
 \smallskip

 \thicksize=0pt
 \hrule \vskip .04in \hrule
\begintable
 Scenario | $b\anti b$ | $t\anti t$ | $\cnone\cnone$ | $\cnone\cntwo$ |
            $\cntwo\cntwo$ | $\cpone\cmone$ | $\hl Z$ \cr
$\dilp_3$ | 0.726 | --- | 0.040 | 0.076 | 0.034 | 0.075 | --- \nr
$\dilm_1$ | 0.113 | --- | 0.009 | 0.144 | 0.504 | 0.189 | 0.031 \nr
$\dilm_4$ | 0.128 | --- | 0.015 | 0.160 | 0.407 | 0.231 | 0.048 \nr
$\dilp_4$ | 0.096 | --- | 0.152 | 0.230 | 0.087 | 0.419 | 0.010 \nr
$\dilp_5$ | 0.240 | --- | 0.076 | 0.218 | 0.153 | 0.286 | 0.008 \nr
$\dilp_7$ | 0.271 | 0.198 | 0.041 | 0.152 | 0.136 | 0.176 | 0.006 \nr
$\sdilm_1$ | 0.819 | 0.009 | 0.005 | 0.028 | 0.038 | 0.037 | 0.001 \nr
$\sdilm_2$ | 0.255 | 0.470 | 0.009 | 0.056 | 0.089 | 0.091 | 0.009 %\nr
\endtable
 \hrule \vskip .04in \hrule
 \vfill
\endinsert

\REF\isasusy{F. Paige and S. Protopopescu, in {\it Supercollider
Physics}, p. 41, ed. D. Soper (World Scientific, 1986);
H. Baer, F. Paige, S. Protopopescu, and X. Tata, in {\it Proceedings of
the Workshop on Physics at Current Accelerators and Supercolliders}, eds. J.
Hewett, A. White, and D. Zeppenfeld, (Argonne National Laboratory, 1993).}
Detailed decay tables for the Higgs bosons and superparticles
were generated using ISASUSY,\refmark\isasusy\
and cross-checked using independent programs
developed for the work of Ref.~[\habgun].
The important Higgs branching ratios as a function of scenario
are presented in Table~\brs.  Note that the cumulative effect of
the SUSY decay modes is generally to substantially reduce the SM
particle modes, unless $\tanb$ is very large (as in the $\dilp_3$ and
$\sdilm_1$ cases)
in which case the $b\anti b$ mode can still be dominant. Especially
dramatic is the dominance of the $\snu\,\snu$ decay modes in the $\dilm_1$
and $\dilm_4$ cases, which has a drastic impact given
that in these cases the $\snu$ itself decays invisibly (see Ref.~[\bgkp]).

The formalism for computing the rate of Higgs boson production in
$\gam\gam$ collisions is well-established.\refmark{\hhg,\habgun}\
An approximate result\foot{In practice, we employ
a more accurate numerical procedure.}
for the number of Higgs bosons produced
at a back-scattered-laser-beam facility is
$$
N(\gam\gam\rta\h)={4\pi^2 \Gamma(\h\rta\gam\gam) \over  \mh^3}
\yh F(\yh) (1+\vevlam_{\yh})
{L}_{\epem}\,,\eqn\nevents$$
where $\yh\equiv \mh/\eepem$, and $F(\yh)$ and $\vevlam_{\yh}$ are obtained
by convoluting together the spectra and polarizations for the back-scattered
photons. In computing $\Gamma(\h\rta\gam\gam)$,
the full set of SUSY and SM particle loops is included.\REF\program{The program
employed was an expanded version of that developed for the work of
Ref.~[\habgun].}\
For each given
scenario these contributions are completely known, since all parameters
and masses of the MSSM are fixed.
In computing $F(\yh)$ we have been as optimistic as possible, choosing
the laser-photon polarizations, $e^+$ and $e^-$ polarizations,
and machine energy so that the $\gam\gam$ spectrum is sharply peaked
and is centered at the Higgs mass of interest. The most highly-peaked spectrum
is obtained by choosing large polarizations
for the $e^+$ and $e^-$ (we adopt $\lam_e=\lam_e^\prime=+0.45$),
large polarizations (opposite those for the $e^+,e^-$)
for the laser photons (we take $P_c=P_c^\prime=-1$), and
as large a value for the $\xi$ parameter as possible (we employ $\xi=4.8$)
without going above pair production threshold.
\foot{We remind the reader that these choices also maximize $1+\vevlam_{\yh}$,
which not only enhances the Higgs boson production rate, but also minimizes
all of the two-body continuum background channels of interest:
$b\anti b$, $t\anti t$, $\cpone\cmone$, and $\slep\ \slep$.}
(For details see Refs.~[\caldwelletal], [\telnovi], and [\telnovii].)
For these choices, the spectrum is peaked in the vicinity of $\yh=0.79$
for which $\yh F(\yh)(1+\vevlam_{\yh})\sim 3.5$,
with $\vevlam_{\yh}\sim 0.94$.  (The corresponding value of $F(\yh)\sim 2.3$
is illustrated, for example, in Fig.~9d of Ref.~[\caldwelletal],
for a very similar back-scattered-laser-beam configuration.)

\TABLE\totalrates{A tabulation of inclusive Higgs boson production
rates as a function of scenario.  We assume $L=10\fbi$ and
have optimized the $\gam\gam$ energy spectrum and collider energy.
The corresponding optimal $\epem$ energy (in GeV) for each Higgs boson
is tabulated.}
 \midinsert
 \titlestyle{\tenpoint
 Table \totalrates: A tabulation of inclusive Higgs boson production
rates as a function of scenario.  We assume $L=10\fbi$ and
have optimized the $\gam\gam$ energy spectrum and collider energy.
The corresponding optimal $\epem$ energy (in GeV) for each Higgs boson
is tabulated.
}
 \smallskip

 \thicksize=0pt
 \hrule \vskip .04in \hrule
 \begintable
 Scenario |  $\mha$ | $\ha$ rate | $\sqrt s_{opt}$ |
             $\mhh$ | $\hh$ rate | $\sqrt s_{opt}$ \cr
$\dilp_3$ | 180 | 56 | 228 | 180 | 40 | 228 \nr
$\dilm_1$ | 190 | 363 | 240 | 205 | 466 | 260 \nr
$\dilm_4$ | 244 | 210 | 309 | 255 | 190 | 323 \nr
$\dilp_4$ | 250 | 70  | 316 | 255 | 46  | 324 \nr
$\dilp_5$ | 300 | 14  | 381 | 302 | 50  | 382 \nr
$\dilp_7$ | 350 | 6   | 443 | 351 | 59  | 445 \nr
$\sdilm_1$ | 357 | 0.5 | 451 | 357 | 11 | 452 \nr
$\sdilm_2$ | 424 | 38 | 538 | 426 | 17 | 538
\endtable
 \hrule \vskip .04in \hrule
\vfill
 \endinsert

The resulting total rates for $\ha$ and $\hh$ production for each
scenario appear in Table~\totalrates\ (assuming an integrated
luminosity of $L\equiv L_{\epem}=10\fbi$, such as might be accumulated
in one year of operation), along with the corresponding choices
of optimal $\sqrt s$ for the $\epem$ collider. Note that the decline
in production rate with increasing Higgs boson mass due to the
$\mh^{-3}$ factor in Eq.~\nevents\ is significantly modulated
by variations in $\Gamma(\h\rta\gam\gam)$, which in particular
is sharply suppressed at large $\tanb$ due to enhanced cancellations
from the $b$-quark loop contribution, whereas it turns out to be
comparatively enhanced for the $\sdilm_2$ scenario.

We recognize that the use of a highly-peaked spectrum for initial
discovery of the Higgs bosons is unrealistic in practice,
as it requires scanning in order to discover a given Higgs boson.
However, we have adopted a highly-peaked spectrum
for two reasons. First, it yields the most optimistic results possible,
which will not prove to be terribly promising. Second, it gives an
accurate representation of what would be possible should the mass of a given
Higgs boson already be known, in which case $\gam\gam$ collision detection
would be a second generation experiment motivated by the importance
of determining $\Gamma(\h\rta\gam\gam)$.
In practice, $\ha$ and $\hh$ Higgs boson {\it searches} in $\gam\gam$
collisions (\ie\ prior to their discovery elsewhere)
would probably employ a fixed $\sqrt s$, in which case
it is probably most reasonable to assume
that $\mha$ and $\mhh$ would not be $\sim 0.79 \sqrt s$.
The above-specified back-scattered laser beam configuration
(for which $F(\yh)$ falls to $\sim 1$ for $\yh\lsim 0.6$) would be employed
in order to explore for Higgs bosons with $\mh\sim 0.6-0.8\sqrt s$,
while the configuration
$\lam_e\sim\lam_e^\prime\sim 0.45$, $P_c\sim P_c^\prime\sim +1$
(for which $F(\yh)$ exhibits a spectrum that is broadly-peaked with
$F(\yh)\sim 1.7$ in the vicinity of $\yh\sim 0.4$ falling below 1 for
$\yh$ below 0.1 and above 0.6 --- see Fig.~9b of Ref.~[\caldwelletal])
would be employed to explore for Higgs bosons
below $0.6\sqrt s$.   Then,
the true rates for the various channels considered here
would most typically be between 20\% and 50\% lower than those quoted
below assuming we sum over two runs with
an integrated luminosity of $L=10\fbi$ in {\it each} of the
two complementary back-scattered laser beam configurations outlined above.

We turn next to rates in specific channels. Tree-level backgrounds
are present for the $b\anti b$, $t\anti t$, $\cpone\cmone$, and $\slep\ \slep$
channels.  The $\cnone\cnone$ channel is invisible, while the $\cnone\cntwo$
and $\cntwo\cntwo$ backgrounds only arise at one loop.  The $\hl\hl$
and $\hl Z$ channels we regard as background free, assuming that the $\hl$
and $Z$ masses can be reconstructed with reasonable accuracy in the $b\anti b
b\anti b$ and $b\anti b Z$ (with $Z$ visible) modes.

\TABLE\bbtt{A tabulation of Higgs boson
signal and background rates (assuming $L=10\fbi$)
for the $b\anti b$ and $t\anti t$ channels. In computing the background rates
a final state mass resolution of 10 GeV is assumed.}
 \midinsert
 \titlestyle{\tenpoint
 Table \bbtt: A tabulation of Higgs boson
signal and background rates (assuming $L=10\fbi$)
for the $b\anti b$ and $t\anti t$ channels. In computing the background rates
a final state mass resolution of 10 GeV is assumed.
}
 \smallskip

 \thicksize=0pt
 \hrule \vskip .04in \hrule
 \begintable
 Scenario | $\ha\rta b\anti b$ | Bkgnd. |
$\hh\rta b\anti b$ | Bkgnd. | $\ha\rta t\anti t$ | Bkgnd. |
$\hh\rta t\anti t$ | Bkgnd. \cr
$\dilp_3$ | 41 | 770 | 31 | 770 | --- | --- | --- | --- \nr
$\dilm_1$ | 41 | 670 | 21 | 570 | --- | --- | --- | --- \nr
$\dilm_4$ | 27 | 320 | 14 | 290 | --- | --- | --- | ---  \nr
$\dilp_4$ | 7 | 300 | 7 | 290 | --- | --- | --- | ---  \nr
$\dilp_5$ | 3 | 180 | 17 | 170 | --- | --- | --- | --- \nr
$\dilp_7$ | 2 | 120 | 27 | 120 | 1 | 350 | 2 | 370 \nr
$\sdilm_1$ | 0.4 | 110 | 9 | 110 | 0.005  | 430 | 0.02 | 430 \nr
$\sdilm_2$ | 10 | 70 |  5 | 77 | 18 | 580 | 5 | 570
\endtable
 \hrule \vskip .04in \hrule
\vfill
 \endinsert

We examine first the $b\anti b $ and $t\anti t$
final state decay modes and their backgrounds.  The rates are summarized in
Table~\bbtt.  In obtaining these rates we have
not included the efficiency penalty that will inevitably
arise in experimentally isolating the $b$ and $t$ final states.
Further, in estimating background rates, we have assumed a 10 GeV mass
resolution, which might be achievable for $b\anti b$ final states
but is certainly far too optimistic for the $t\anti t$ channel.
Even with these optimistic procedures, discovery
of the $\hh$ and $\ha$ appears quite difficult.
The statistical significance, $\nsd\equiv S/\sqrt B$,
achieved by combining the $\ha$ and $\hh$
signals (not really allowed in cases where the
$\sqrt s$ values needed to achieve the optimal rates are somewhat different)
and using the average of the two backgrounds is always below $\nsd=3$, and
declines to no more than $\nsd=1$ or 2 at higher Higgs masses. Thus, even for
our optimal $\gam\gam$ spectrum and resolution choices, roughly $L\gsim 60\fbi$
would be required for these channels to provide viable signals for
most scenarios.

\TABLE\hlhlhlz{A tabulation of signal rates (assuming $L=10\fbi$)
in the $\hh\rta\hl\hl$ and $\ha\rta \hl Z$ channels.}
 \midinsert
 \titlestyle{\tenpoint
 Table \hlhlhlz: A tabulation of signal rates (assuming $L=10\fbi$)
in the $\hh\rta\hl\hl$ and $\ha\rta \hl Z$ channels.
}
 \smallskip

 \thicksize=0pt
 \hrule \vskip .04in \hrule
 \begintable
 Scenario | $\ha\rta \hl Z$ rate | $\hh\rta \hl\hl$ rate \cr
$\dilp_3$ | --- | --- \nr
$\dilm_1$ | 11 | 48 \nr
$\dilm_4$ | 10 | 28 \nr
$\dilp_4$ | 0.7 | 2.9 \nr
$\dilp_5$ | 0.1 | 1.4 \nr
$\dilp_7$ | 0.04 | 0.9 \nr
$\sdilm_1$ | 0.0005 | 0.14 \nr
$\sdilm_2$ | 0.35 | 1.4
\endtable
 \hrule \vskip .04in \hrule
\vfill
 \endinsert

Let us next examine the $\ha\rta \hl Z$ and $\hh\rta\hl\hl$ channels. Raw event
rates are presented in Table~\hlhlhlz.  We see immediately that these channels
only show a reasonable level of promise in the case of the $\dilm_1$ and
$\dilm_4$ scenarios. These two scenarios
illustrate more generally the ingredients required
in order that the $\hl Z$ and $\hl\hl$ channels yield viable discovery signals:
i) the $\ha$ and $\hh$ masses are sufficiently modest that the
$\mh^{-3}$ factor in Eq.~\nevents\ does not yield too much rate suppression,
but sufficiently large that $\hl Z$
and $\hl\hl$ decays are kinematically allowed;
ii) the value of $\tanb$ is moderate so that the $b\anti b$ decay channel of
the Higgs bosons does not overwhelm all others and the $b$-quark loop is
not enhanced so as to cause cancellations
that yield small values for $\Gamma(\ha,\hh \rta\gam\gam)$; and
iii) the Higgs masses are small enough that SUSY decay modes still
suffer some kinematical suppression.
Of course, in realistically assessing the visibility of the $\hl Z$ and
$\hl\hl$ signals one must take into account
the fact that $\hl\hl\rta b\anti b b \anti b$ and $\hl Z\rta b\anti b +visible$
branching fractions [typically $BR(\hl\rta b\anti b)\sim 0.9$ and
$BR(Z\rta visible)\sim 0.8$]
will reduce the effective rates for useful channels
and the fact that to isolate these channels from QCD backgrounds
it will be necessary to tag
at least one of the $b$-quark jets (with roughly 60\% efficiency).
Consequently, the effective
rates for these promising channels will be somewhat marginal even in the
most favorable scenarios, unless $L>10\fbi$ is accumulated.

\TABLE\cnmoderates{A tabulation of signal rates (assuming $L=10\fbi$)
in the $\cnone\cntwo$
and $\cntwo\cntwo$ final states, before including $\cntwo$
decay branching fractions.}
 \midinsert
 \titlestyle{\tenpoint
 Table \cnmoderates: A tabulation of signal rates (assuming $L=10\fbi$)
in the $\cnone\cntwo$
and $\cntwo\cntwo$ final states, before including $\cntwo$
decay branching fractions.
}
 \smallskip

 \thicksize=0pt
 \hrule \vskip .04in \hrule
 \begintable
 Scenario | $\ha\rta \cnone\cntwo$ | $\ha\rta \cntwo\cntwo$ |
            $\hh\rta \cnone\cntwo$ | $\hh\rta \cntwo\cntwo$ \cr
$\dilp_3$ | 4 | 2 | 2 | 0.4 \nr
$\dilm_1$ | 52 | 183 | 14 | 41 \nr
$\dilm_4$ | 34 | 86 | 10 | 20 \nr
$\dilp_4$ | 16 | 6 | 6 | 2  \nr
$\dilp_5$ | 3 | 2 | 7 | 3  \nr
$\dilp_7$ | 1 | 1 | 7 | 3   \nr
$\sdilm_1$ | 0.01 | 0.02 | 0.2 | 0.2   \nr
$\sdilm_2$ | 2 | 3 | 1 | 1
\endtable
 \hrule \vskip .04in \hrule
\vfill
 \endinsert

\TABLE\cndecays{A tabulation of branching ratios ($BR$)
for the three basic $\cntwo$ decay channels.}
 \midinsert
 \titlestyle{\tenpoint
 Table \cndecays: A tabulation of branching ratios ($BR$)
for the three basic $\cntwo$ decay channels.
}
 \smallskip

 \thicksize=0pt
 \hrule \vskip .04in \hrule
 \begintable
 Scenario | $BR(ll+\etmiss)$ | $BR(jj+\etmiss)$ | $BR(\etmiss)$ \cr
$\dilp_3$ | 0.082 | 0.067 | 0.851 \nr
$\dilm_1$ | 0.017 | 0.006 | 0.977  \nr
$\dilm_4$ | 0.027 | 0.014 | 0.959 \nr
$\dilp_4$ | 0.301 | 0.187 | 0.510 \nr
$\dilp_5$ | 0.314 | 0.205 | 0.481 \nr
$\dilp_7$ | 0.266 | 0.204 | 0.530 \nr
$\sdilm_1$ | 0.206 | 0.362 | 0.432 \nr
$\sdilm_2$ | 0.251 | 0.355 | 0.394
\endtable
 \hrule \vskip .04in \hrule
\vfill
 \endinsert

Could SUSY decay channels save the day?  Let us first focus on
the tree-level-background-free $\cnone\cntwo$ and $\cntwo\cntwo$
channels. The rates for these channels for the $\ha$ and $\hh$
are given in Table~\cnmoderates.
In order to assess the possible
utility of these rates we need to include the $\cntwo$ decays.
The primary decays of the $\cntwo$
are of three basic types: $ll+\etmiss$ (often via the two-body $l\slepr$ mode,
with $\slepr\rta l\cnone$), $jj+\etmiss$ (in which we
include $\tau\tau+\etmiss$, aside from which it is always a three-body decay),
and pure $\etmiss$ (often via two-body $\snu\nu$ modes where the $\snu$
decays invisibly via $\snu\rta \nu\cnone$).
The branching ratios for these three basic types of $\cntwo$ decay are given
in Table~\cndecays\ as a function of scenario.

\TABLE\cnrates{A tabulation of rates (assuming $L=10\fbi$)
for the five classes of visible $\cnone\cntwo+\cntwo\cntwo$
final state after combining $\ha$ and $\hh$ production.}
 \midinsert
 \titlestyle{\tenpoint
 Table \cnrates: A tabulation of rates (assuming $L=10\fbi$)
for the five classes of visible $\cnone\cntwo+\cntwo\cntwo$
final state after combining $\ha$ and $\hh$ production.
}
 \smallskip

 \thicksize=0pt
 \hrule \vskip .04in \hrule
 \begintable
 Scenario | $ll+\etmiss$ | $jj+\etmiss$ | $ll+jj+\etmiss$
          | $ll+ll+\etmiss$ | $jj+jj+\etmiss$ \cr
$\dilp_3$ | 0.8 | 0.7 | 0.03 | 0.02 | 0.01 \nr
$\dilm_1$ | 8 | 3 | 0.05 | 0.06 | 0.008 \nr
$\dilm_4$ | 7 | 3 | 0.08 | 0.08 | 0.02 \nr
$\dilp_4$ | 9 | 6 | 0.9 | 0.7 | 0.3 \nr
$\dilp_5$ | 5 | 3 | 0.6 | 0.5 | 0.2 \nr
$\dilp_7$ | 3 | 3 | 0.4 | 0.3 | 0.2 \nr
$\sdilm_1$ | 0.1 | 0.1 | 0.03 | 0.009 | 0.03 \nr
$\sdilm_2$ | 2 | 2 | 0.7 | 0.3 | 0.5
\endtable
 \hrule \vskip .04in \hrule
\vfill
 \endinsert

The types of Higgs boson final states that result are of six basic classes.
The $\cnone\cntwo$ decay mode of the $\ha$ and $\hh$ can lead to
a purely invisible decay channel, which we discard as unusable,
a channel with two leptons and missing energy,
$ll+\etmiss$ (where both $l$'s come from the $\cntwo$),
and a channel with two jets and missing energy,
$jj+\etmiss$ (where we include $\tau$ leptons in the $j$).
The $\cntwo\cntwo$ decay mode can lead to these same final states
and, in addition,
a two-lepton-two-jet plus missing energy final state, $ll+jj+\etmiss$,
a four-lepton plus missing energy final state, $ll+ll+\etmiss$,
and a four-jet plus missing energy final state, $jj+jj+\etmiss$.
In computing the rates for these final states we combine the
events coming from the $\ha$ and $\hh$  --- these have similar
mass, and mass reconstruction in the final state is not possible due
to the missing energy content.  The resulting event rates for
each class of final state are displayed in Table~\cnrates.

We see that only the $ll+\etmiss$ and $jj+\etmiss$ channels have
a non-negligible number of events, and that even these rates are
very modest.  The reasons for this are several, and can be traced
from Tables~\cnmoderates\ and \cndecays. For the $\dilm_1$ and $\dilm_4$
scenarios, Higgs boson production rates were high, but decays for the $\cntwo$
are completely dominated by totally invisible channels.
For the other scenarios, visible $\cntwo$ decays have a substantial
branching fraction but Higgs boson production rates are much more modest.
We cannot say if this conspiracy is a general phenomenon, or
simply specific to the dilaton-like boundary conditions employed here.

Are the $ll+\etmiss$ and/or $jj+\etmiss$
events sufficiently unique to provide a viable signal?
We are pessimistic in this regard, since many large rate processes
can potentially yield backgrounds. Consider first the $ll+\etmiss$ channel.
We shall see that
tree-level $\cpone\cmone$ continuum production has a {\it very} high rate,
and since the $\cpmone$ have a significant branching fraction to
$l+\etmiss$, we will have a large number of $ll+\etmiss$
final states from this source.  Even though the two leptons of a signal event
both derive from a single $\cntwo$, they will not tend to be
terribly well-collimated due to the large role played by the $\etmiss$
component of a given $\cntwo$ decay.  Thus, we believe (but we
have not performed a Monte Carlo study) that event topology will not
allow a sufficiently efficient means of discriminating the signal
of interest from this very large background. In addition,
$\slep\ \slep$ production also has a very high rate and also contributes
to the $ll+\etmiss$ channel.
Regarding the $jj+\etmiss$ channel, once again $\cpone\cmone$
will yield a background when one $\cpmone$ decays hadronically to two jets plus
missing energy and the other decays leptonically and the lepton is `missed'.
In addition, $\gam\gam\rta jet+jet$ rates are very high, and will
inevitably have a significant detector-dependent missing energy tail.
SUSY production processes can also contribute backgrounds; for example,
$\gam\gam\rta \sq\,\sq$ contributes when both squarks decay to $q\cnone$.
Thus, even before inclusion of detection efficiencies,
we are relatively certain that the low Higgs boson signal event rates would not
constitute viable signals. (Detailed studies will not be pursued here.)
Models with very different boundary conditions could perhaps
yield more viable Higgs boson signal rates in these channels.

\TABLE\cpslepmodes{A tabulation of signal rates (assuming $L=10\fbi$)
in the $\cpone\cmone$
and $\slep\ \slep$ final states. Backgrounds in these channels are
also given for the (unrealistically small) final state mass
resolution of 10 GeV.}
 \midinsert
 \titlestyle{\tenpoint
 Table \cpslepmodes: A tabulation of signal rates (assuming $L=10\fbi$)
in the $\cpone\cmone$
and $\slep\ \slep$ final states. Backgrounds in these channels are
also given for the (unrealistically small) final state mass
resolution of 10 GeV.
}
 \smallskip

 \thicksize=0pt
 \hrule \vskip .04in \hrule
 \begintable
 Scenario | $\ha\rta \cpone\cmone$ | Bkgnd. |
            $\hh\rta \cpone\cmone$ | Bkgnd. | $\hh\rta\slep\  \slep$ |
            Bkgnd. \cr
$\dilp_3$ | 4 | 13000 | 3 | 13000 | .01 | 4400 \nr
$\dilm_1$ | 69 | 7900 | 52 | 8100 | 51 | 6800 \nr
$\dilm_4$ | 49 | 4800 | 29 | 4600 | 15 | 3600 \nr
$\dilp_4$ | 29 | 6600 | 13 | 6400 | 2 | 3800 \nr
$\dilp_5$ | 4 | 3300 | 12 | 3300 | 0.7 | 2200 \nr
$\dilp_7$ | 1 | 1900 | 11 | 1900  | 0.6 | 1400 \nr
$\sdilm_1$ | 0.02 | 1800 | 0.5 | 1800 | 0.001 | 780 \nr
$\sdilm_2$ | 3 | 1200 | 2 | 1200 | 0.07 | 780
\endtable
 \hrule \vskip .04in \hrule
\vfill
 \endinsert

The remaining SUSY-channel possibilities are the $\cpone\cmone$ and
$\slep\ \slep$ channels.  Generally speaking, both primarily yield $ll+\etmiss$
final states (although the $\cpone$ can decay also to jets, this mode
is generally smaller than the leptonic mode).  So in some sense
these channels should be considered together
and also combined (to the extent that the topologies
do not differ much) with the $ll+\etmiss$ events deriving
from the $\cnone\cntwo$ and $\cntwo\cntwo$ decay channels.
(Of course, in the latter case the two $l$'s must be of the same type
whereas for the $\cpone\cmone$ modes they can be of different types.)
For purposes of discussion, we shall keep all these different channels
separate. The event rates for these channels
are given in Table~\cpslepmodes, along with the direct tree-level backgrounds
assuming a final state mass resolution of 10 GeV. Such a small resolution
is undoubtedly highly unrealistic given that the $\cpone\cmone$ and
$\slep\ \slep$ final states contain significant missing energy.
A cursory survey of the numbers reveals the impossibility of overcoming
the backgrounds.  (A number of distributions for final leptons
were examined to see if any dramatic increases of $S/B$ could be
achieved by appropriate cuts, but no effective cuts were found.)
Even if we ignore all topology differences and add in the $\cnone\cntwo$
and $\cntwo\cntwo$ events of the $ll+\etmiss$ type, the signal rates
remain very small compared to the backgrounds.

\vskip .15in
\noindent{\bf 3. Conclusions}
\vskip .075in

We are forced to conclude that detection of the $\hh$ and $\ha$
in $\gam\gam$ collisions at a back-scattered laser beam facility
could prove extremely difficult in models where
SUSY decays of the Higgs bosons are significant, unless
integrated luminosities much higher than $L=10\fbi$ could be provided.
For the models explored here we found that,
even for a completely optimized $\gam\gam$ energy spectrum,
for $L=10\fbi$ the $b\anti b$ and $t\anti t$
channel rates are generally reduced to too low a level relative
to the corresponding continuum backgrounds to provide a viable Higgs boson
signal. The SUSY decay modes themselves do not appear to
have large enough rates relative to expected backgrounds.
The only channels that have a significant chance
of revealing a signal are the (background free) $\ha\rta \hl Z\rta b\anti b
Z_{vis}$ and $\hh\rta \hl\hl\rta b\anti b b\anti b$ modes,
and even the most promising specific scenarios
that we have examined yield only very modest event rates despite
the optimization of the $\gam\gam$ spectrum. Considering all
possible channels, for most of the scenarios examined here
$L\gsim 50\fbi$ would be needed in order to obtain
at least one viable signal.

The basic problem is that once SUSY decay modes are allowed, the
large number of decay channels means that no single decay channel is
likely to be dominant (with the exception of the largely or
completely invisible $\snu\,\snu$ channel). Consequently, no single final state
mode obtains a high event rate. The only exception to this rule arises
if $\tanb$ is large, in which case the $b\anti b$ decay mode
is dominant for both the $\hh$ and $\ha$, and the only issue
is the absolute production rate of the Higgs bosons themselves.
Unfortunately, as noted previously in Ref.~[\habgun],\refmark\habgun\
for Higgs boson masses in the 200--500 GeV range there is a general tendency
for the enhanced $b$-quark loop to significantly cancel against
other loops contributing to the one-loop $\gam\gam$ couplings of the $\ha$
and $\hh$, thereby leading to suppressed production rates. (Compare
the rates of the high-$\tanb$ scenarios, $\dilp_3$ and $\sdilm_1$, in
Table~\totalrates\ to those for lower $\tanb$ scenarios with similar $\mha$.)

Of course, there are certainly SUSY scenarios that will yield viable
$\ha$ and $\hh$ signals in the $b\anti b$, $t\anti t$, $\hl Z$,
and $\hl\hl$ modes,
in particular any model in which all SUSY states are more massive
than one-half the Higgs boson masses. Nonetheless, we cannot ignore
the fact that the very attractive dilaton-like boundary conditions suggested
by superstring theory generally yield a sufficiently complex
array of $\ha$ and $\hh$ decays as to make their detection in $\gam\gam$
collisions highly problematical.

We conclude that one should not count on being able to see the
$\hh$ and $\ha$ in $\gam\gam$ collisions for integrated luminosities
of order $L=10\fbi$ unless we become convinced by other experiments
that the SUSY mass scale is quite high.
This places increased onus on achieving much higher $L$
or on building a machine with $\sqrt s$ sufficiently large that
$\hh\ha$ and $\hp\hm$ pair production will be possible via direct
$\epem$ collisions. With regard to the latter,
the gauge-coupling-unified models typified by
those explored here suggest that $\sqrt s$ above 500 GeV is generally
required, with 1 TeV providing adequate energy for a large section
of model parameter space.  Of course, it remains to explore
the degree to which SUSY decay modes and backgrounds complicate the detection
of the above pair states.\Ref\inprogress{Work in progress.}\

\vskip .15in
\noindent{\bf Acknowledgements}
\vskip .075in

This work was partially supported by the Department of Energy.
JFG thanks the Aspen Institute for Physics for support and hospitality
during the final stages of this project. We wish to acknowledge the
contributions of H. Baer, H. Haber, and H. Pois to the MSSM scenarios
and/or programs employed.

\refout
%\tabout %%%%%%%%%%%%physrev -- include
\end
\bye